\documentclass[aps,prb,preprint,groupedaddress, showpacs, showkeys]{revtex4}

\usepackage{subfigure}
\usepackage{graphicx}

\newcommand{\STO}{SrTiO$_3$}
\newcommand{\LAO}{LaAlO$_3$}
\newcommand{\LNO}{LaNiO$_3$}
\newcommand{\LaNiO}{La$_2$NiO$_4$}
\newcommand{\LTO}{LaTiO$_3$}

\newcommand{\Tg}{t$_{2g}$}
\newcommand{\Eg}{e$_g$}
\newcommand{\Lth}{$L_3$}
\newcommand{\Ltw}{$L_2$}
\newcommand{\TL}{Ti $L$-edge}
\newcommand{\NL}{Ni $L$-edge}
\newcommand{\OK}{O $K$-edge}
\newcommand{\Ti}{Ti$^{3+}$}
\newcommand{\Ni}{Ni$^{3+}$}

\begin{document}

\title{Experimental verification of orbital engineering at the atomic scale: charge transfer and symmetry breaking in nickelate heterostructures}

\author{Patrick J. Phillips}
\affiliation{Nanoscale Physics Group, University of Illinois at Chicago, Chicago, IL 60607}
\author{Paolo Longo}
\affiliation{Gatan Inc., Pleasanton, CA 94588}
\author{Alexandru B. Georgescu}
\affiliation{Department of Physics, Department of Applied Physics, and Center for Research on Interface Structures and Phenomena (CRISP), Yale University, New Haven, CT 06511}
\author{Eiji Okunishi}
\affiliation{JEOL Ltd., Tokyo 196-8558, JAPAN }
\author{Xue Rui}
\affiliation{Nanoscale Physics Group, University of Illinois at Chicago, Chicago, IL 60607}
\author{Ankit S. Disa}
\affiliation{Department of Physics, Department of Applied Physics, and Center for Research on Interface Structures and Phenomena (CRISP), Yale University, New Haven, CT 06511}
\author{Fred Walker}
\affiliation{Department of Physics, Department of Applied Physics, and Center for Research on Interface Structures and Phenomena (CRISP), Yale University, New Haven, CT 06511}
\author{Sohrab Ismail-Beigi}
\affiliation{Department of Physics, Department of Applied Physics, and Center for Research on Interface Structures and Phenomena (CRISP), Yale University, New Haven, CT 06511}
\author{Charles H. Ahn}
\affiliation{Department of Physics, Department of Applied Physics, and Center for Research on Interface Structures and Phenomena (CRISP), Yale University, New Haven, CT 06511}
\author{Robert F. Klie}
\affiliation{Nanoscale Physics Group, University of Illinois at Chicago, Chicago, IL 60607}

\date{\today}

\begin{abstract}
Epitaxial strain, layer confinement and inversion symmetry breaking have emerged as powerful new approaches to control the electronic and atomic-scale structural properties in complex metal oxides. Nickelate heterostructures, based on RENiO$_3$, where RE is a trivalent rare-earth cation, have been shown to be relevant model systems since the orbital occupancy, degeneracy, and, consequently, the electronic/magnetic properties can be altered as a function of epitaxial strain, layer thickness and superlattice structure. One such recent example is the tri-component \LTO-\LNO-\LAO\ superlattice, which exhibits charge transfer and orbital polarization as the result of its interfacial dipole electric field. A crucial step towards control of these parameters for future electronic and magnetic device applications is to develop an understanding of both the magnitude and range of the octahedral network’s response towards interfacial strain and electric fields. An approach that provides atomic-scale resolution and sensitivity towards the local octahedral distortions and orbital occupancy is therefore required. Here, we employ atomic-resolution imaging coupled with electron spectroscopies and first principles theory to examine the role of interfacial charge transfer and symmetry breaking in a tricomponent nickelate superlattice system.  We find that nearly complete charge transfer occurs between the \LTO\ and \LNO\ layers, resulting in a Ni$^{2+}$ valence state. We further demonstrate that this charge transfer is highly localized with a range of about 1 unit cell, within the \LNO\ layers. The results presented here provide important feedback to synthesis efforts aimed at stabilizing new electronic phases that are not accessible by conventional bulk or epitaxial film approaches.
\end{abstract}


\maketitle

\section{Introduction}
For many technologically-relevant materials systems, in particular transition-metal oxides (TMOs), the orbital structure (relative energies, filling, etc.) directly correlates to the material's resulting properties.\cite{Boris11, Chakhalian11, Frano13, Kumah14, Scherwitzl11} For example, systems such as the manganites (colossal magnetoresistance),\cite{Tokura99} the cobaltates (spin-state transitions),\cite{Maris03, Klie07} and the cuprates (high-temperature superconductivity)\cite{Feiner92, Pavarini01} owe their behaviors to specific configurations of the electronically active transition-metal cation $d$ orbitals, which, for near-cubic symmetry, are split into the (lower energy) \Tg\ and (higher energy) \Eg\ orbitals. The development of atomically precise growth techniques for oxides has opened up the possibility of controlling orbital configurations via layered heterostructures.

\Ni\ ($d^7$) is a $d$ orbital  open-shell system, with fully occupied \Tg\ orbitals and a single electron occupying the twofold-degenerate \Eg\ orbital. \LNO\ (LNO), possessing a pseudocubic perovskite structure, is a material recently explored in the context of orbital engineering, with the goal of breaking its orbital degeneracy and emulating the single-band structure of the cuprates.\cite{Chaloupka08, Hansmann09, Disa15APLMat} A recent publication on a \LTO-\LNO-\LAO\ (LTNAO) superlattice demonstrated the successful breaking of this orbital degeneracy by using atomic-layer synthesis to alter its symmetry and filling; an approximately 50\% change in the occupation of the Ni $d$ orbitals was reported\cite{Disa15PRL} and verified via X-ray absorption spectroscopy and ab initio theory, confirming the creation of an electronic configuration which approaches a single-band Fermi surface.

The three-component superlattice, where 1 unit cell (uc) of \LTO\ (LTO) and  2 uc of LNO are sandwiched between 3 uc of \LAO\ (LAO) develops a large orbital polarization as a result of an inherent inversion symmetry breaking, internal charge transfer, and the resultant ionic polarization.\cite{Chen13, Disa15PRL} The principle is based on the transfer of a single electron from the LTO layer (\Ti) to the LNO layer (\Ni) due to the mismatch in electronegativity of the two ions. The electron transfer creates a charge imbalance and, hence, a dipole field which leads to large polar distortions with polarization pointing towards the NiO$_2$ layer of the LNO. The combination of these polar distortions and the symmetry breaking of the superlattice about the LNO results in asymmetric stretching of the NiO$_6$ oxygen octahedra, leading to a large crystal field splitting and a polarization in the orbital occupations which resembles the arrangement in the high-temperature superconducting cuprates. A previous study\cite{Disa15PRL} infers the charge transfer from the spatially-averaged X-ray absorption spectroscopy (XAS) measurements on the \TL and \NL.

In this work, we focus on a specific superlattice system, consisting of   1 uc of \LTO\ (LTO),  2 uc of LNO and 3 uc of \LAO\ (LAO).   We aim to directly quantify the proposed charge transfer mechanism and determine its range using  this superlattice structure. More specifically, we utilize aberration-corrected scanning transmission electron microscopy (STEM) coupled with both energy dispersive X-ray (EDX) and electron energy loss (EEL) spectroscopies to quantify the orbital manipulation in a nickelate heterostructure; specifically, charge transfer and symmetry breaking at the atomic scale. We directly map the charge transfer with STEM EELS/EDX, providing direct evidence for the key driving force of orbital polarization in the three-component system. Furthermore, we detect the signatures of orbital polarization in this LTNAO superlattice with atomic resolution, as previously suggested by sample-averaged experiments.
 Additionally, we perform first principles density functional theory (DFT) simulations within the local density approximation (LDA) to simulate and verify basic aspects of the electronic structure of these heterostructures using a $c\left(2\times2\right)$ interfacial unit cell that includes rotations and tilts of oxygen octahedra.\cite{Hohenberg64, Kohn65, Perdew81}

It should be noted here that, until recent instrumentation and software advances, an analysis with the spatial and chemical resolution as presented here would not be have been possible, since it requires an imaging probe which has a high enough current density  to generate appreciable X-rays, yet is also able to achieve sub-\AA\ resolution. Indeed, electron microscopy has a rich history in the advanced characterization of oxides, for example, in the atomic-scale imaging of composition, bonding, electron spatial distribution at interfaces, valence determination, etc.\cite{Ohtomo02, Kourkoutis10, Muller08, Gulec15, Varela09, Laffont10, Gauquelin14, Kinyanjui14} By taking advantage of the numerous imaging and spectroscopy modes on advanced aberration-corrected instruments, it is feasible to locally conduct a complete chemical, structural, and electronic characterization at the atomic scale. The high/low angle annular dark field (H/LAADF) and annular bright field (ABF) STEM signals can be simultaneously acquired, resulting in images which are sensitive to atomic number, strain, and light element contrast, respectively.\cite{Kirkland87, Loane92, Phillips12, Findlay10} In terms of spectroscopy, both EDX and EEL signals can be simultaneously acquired, thereby providing atomically-resolved chemical and electronic information.

\section{Methods}
\subsection{Thin Film Synthesis}
The tricomponent superlattice is grown on \LAO\ (001) single crystal substrates using oxygen plasma assisted molecular beam epitaxy. The layering sequence for superlattice is [(\LTO)$_1$-(\LNO)$_2$-(\LAO)$_3$]$\times$ 12 with a total film thickness of $\approx$30 nm. Each layer is grown via co-deposition of the respective elements. The growth is monitored in situ by reflection high energy electron diffraction (RHEED). Post-growth RHEED images display sharp narrow streaks indicative of coherent epitaxy. Ex situ atomic force microscopy reveals low surface roughness ($\approx$1-2 \AA) and unit cell high ($\approx$4 \AA) steps post-growth.  More details on the thin film growth can be found in Ref. \onlinecite{Disa15PRL}

\subsection{X-ray Absorption Spectroscopy}
 XAS measurements shown in Figures \ref{Fig:Ti-XAS} and \ref{Fig:Ok-XAS} were carried out at beamline $U4B$ at the National Synchrotron Light Source. Spectra were recorded in total electron yield mode and normalized by the incident flux as measured by an upstream Au mesh. The energy of the Ti $L$- and O $K$-edges were calibrated with reference to a simultaneously measured TiO$_2$ powder. A linear background is subtracted from the data by fitting to the pre-edge region $\approx$ 5-10 eV below the edge.

\subsection{First-Principles Modeling}
We performed first principles calculations using Density Functional Theory (DFT)\cite{Hohenberg64, Kohn65} with ultrasoft pseudopotentials.\cite{Vanderbilt90, Laasonen91, Laasonen93} To approximate the effects of exchange-correlation, we used the local density approximation (LDA)\cite{Perdew81} as it has been proven to be the best approach for describing bulk LNO bulk from first principles.\cite{Gou11} $k$-point sampling of the Brillouin zone employed a mesh equivalent to a $12\times12\times12$ mesh for a 5-atom pseudocubic bulk unit cell. Band occupations were Gaussian broadened with width 0.03 eV. The plane wave cutoff was 35 Ryd for the wave functions and 280 Ryd for the electron density. Structural relaxations were terminated when all components of atomic forces are below 0.03 eV/\AA in magnitude. The simulated superlattices were periodic in all directions and biaxially strained to the theoretically computed pseudocubic lattice parameter of \LAO\ at 3.71 \AA. Superlattices with $c\left(2\times2\right)$ interfacial unit cells were simulated allowing for octahedral rotations and tilts.

\subsection{Scanning Transmission Electron Microscopy}
Combined atomic EELS and EDS data were acquired using a cold-field emission gun JEOL GrandARM 60 – 300 kV, operated at 160 kV with a beam current of about 85 pA. The microscope is equipped with dual large solid angle SDD detectors for the acquisition of EDS data and a Gatan GIF Quantum ER for the acquisition of EELS data. EELS data was acquired in DualEELS mode where both the the low- and core-loss spectra were acquired  simultaneously. The zero-loss peak, present in the low-loss spectra, can be used to correct and remove all the effects of energy drift allowing a more accurate measurement of any chemical shift. The EELS spectrometer was setup with a dispersion of 0.1 eV / channel  resulting in an energy resolution of 0.5 eV that was needed in order to resolve all the spectral features present in the EELS spectrum moving across the super lattice layers. For simultaneous high-angle annular dark field (HAADF) and annular bright field (ABF) imaging, a probe-convergence angle of 25 mrad was used with a inner detector angle and angular range of 12 mrad for ABF and an inner detector angle of 55 mrad for HAADF imaging.

\section{Results and Discussion}
\begin{figure}[h]
\begin{center}
\includegraphics[width=0.75\columnwidth]{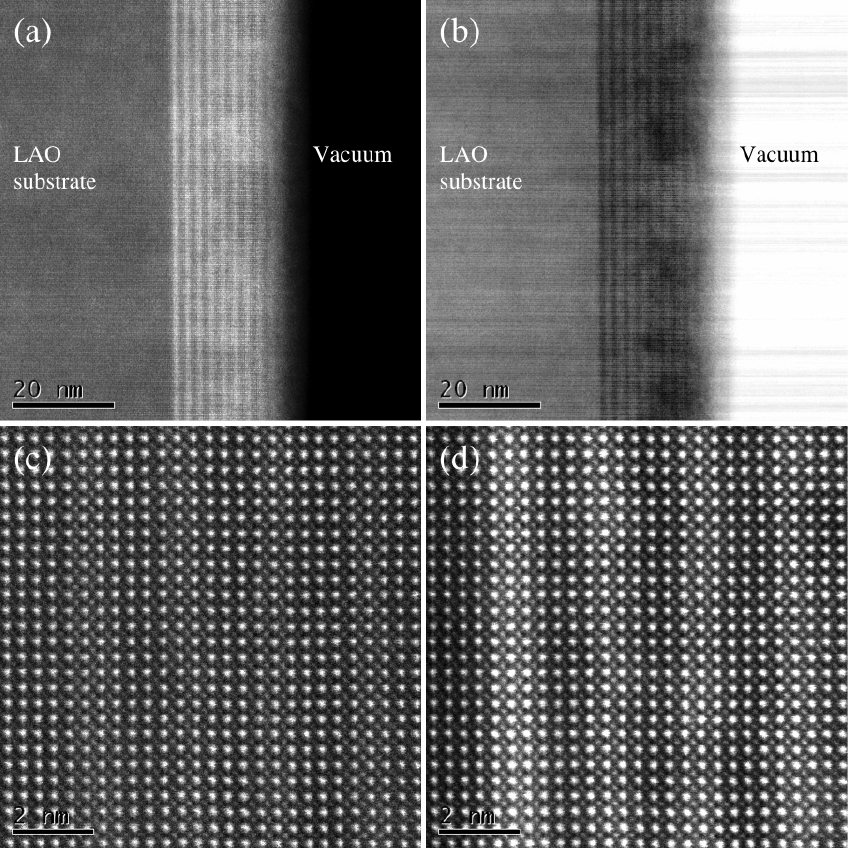}
\caption{(a/b) Low magnification LAADF/ABF image pair of the superlattice structure, with the film growth direction to the right. Some strain is obvious in structure, and the occasional dislocation is observed. These areas were avoided for all chemical and electronic analysis; (c,d) Higher magnification HAADF/LAADF images showing the (lack of) Z-contrast and again, the strain contrast. } \label{Fig:STEM}
\end{center}

\end{figure}

Figure \ref{Fig:STEM} presents a STEM overview of the superlattice structure, following a focused ion beam (FIB) preparation; note that not all the superlattice repeats are visible, as the topmost section of the sample has been milled away in order to render the remainder of the sample sufficiently thin for STEM analysis. Figures \ref{Fig:STEM}a,b) provide a low magnification view of the structure in both LAADF and ABF modes, which reveal some regions of localized strain (anomalously bright/dark in LAADF/ABF), likely from the presence of occasional dislocations; this localized strain is in addition to the strain associated with the superlattice, evidenced by the layering in both the LAADF and ABF images. In general, it is very difficult to discern the identity of the individual layers exclusively via imaging, as there is very little $Z$-contrast gradient across the interfaces of LAO-LTO-LNO (Figure \ref{Fig:STEM}c); thus, chemical spectroscopy is required. Indeed, as LAADF/ABF are considerably less sensitive to atomic number contrast, there is no reason other than strain (the forced constraint on the lattice parameters) for the obvious contrast between layers in Figures \ref{Fig:STEM}a,b,d).

\begin{figure}[h]
\subfigure[]
{
\includegraphics[width=0.465\columnwidth]{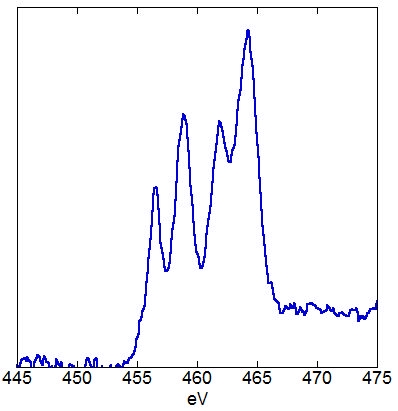}
\label{Fig:Ti-EELS}
}
\subfigure[]
{
\includegraphics[width=0.465\columnwidth]{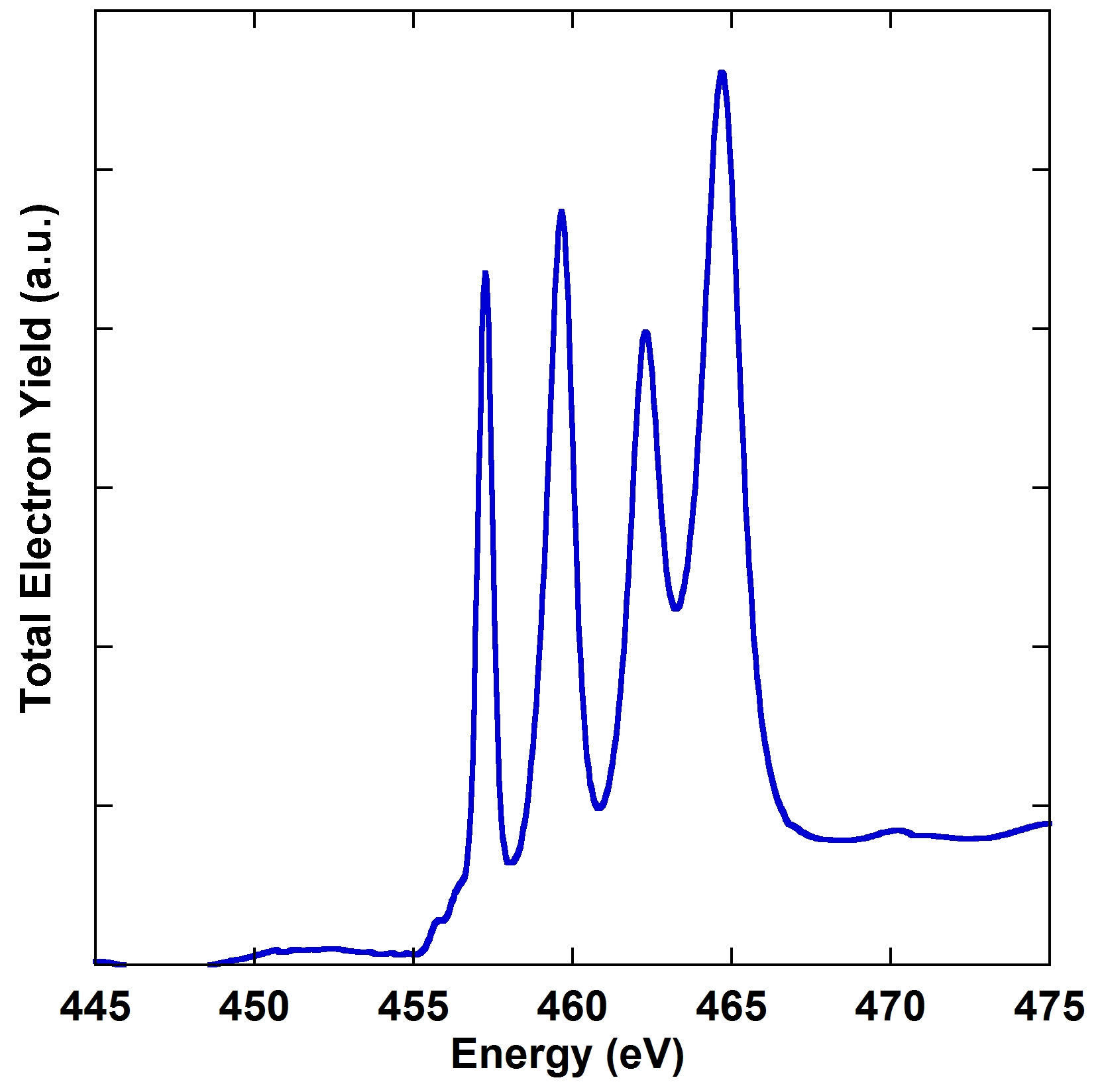}
\label{Fig:Ti-XAS}
}
\subfigure[]
{
\includegraphics[width=0.465\columnwidth]{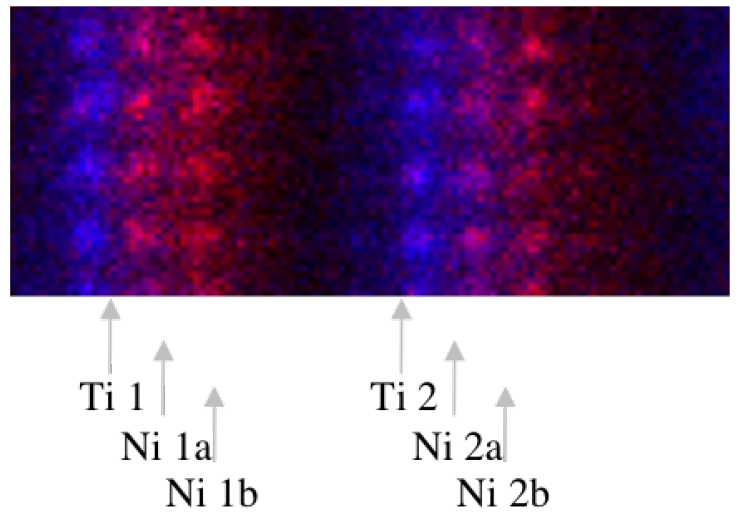}
\label{Fig:Ti-SI}
}
\caption{  a) Electron energy-loss and b) X-ray absorption (right) spectra for the \TL. In the case of EELS, the spectrum was acquired by integrating (vertically) column $Ti$ 2 as shown in c). c) Atomic-resolution spectrum image showing the transition metal oxide atomic columns in the \LTO\ (blue) and \LNO\ (red) layers. } \label{Fig:2}

\end{figure}

Chemical and electronic analyses have made great strides in recent years, owing largely to high-area silicon drift EDX detectors and high-speed/high-sensitivity EEL spectrometers.\cite{Allen12}\cite{Phillips14} Simultaneous acquisition of both signals allows one to avoid the high energy edges in EELS (La, Ni, and Al in this case) in favor of a higher energy dispersion, and to rely on EDX to identify the remaining elements. The higher energy dispersion in EELS then enables the detailed near-edge fine structure analysis of relevant energy loss, including the Ti $L$- and \OK. Several integrated signals are presented in Figure \ref{Fig:2}, coming from both EELS and EDX.  The \OK\ is normalized to the La signal, which is expected to remain relatively constant. Finally, RGB images of various combinations detail the chemical makeup of the LTNAO superlattice structure.

\begin{figure}[h]
\subfigure[]
{
\includegraphics[width=0.465\columnwidth]{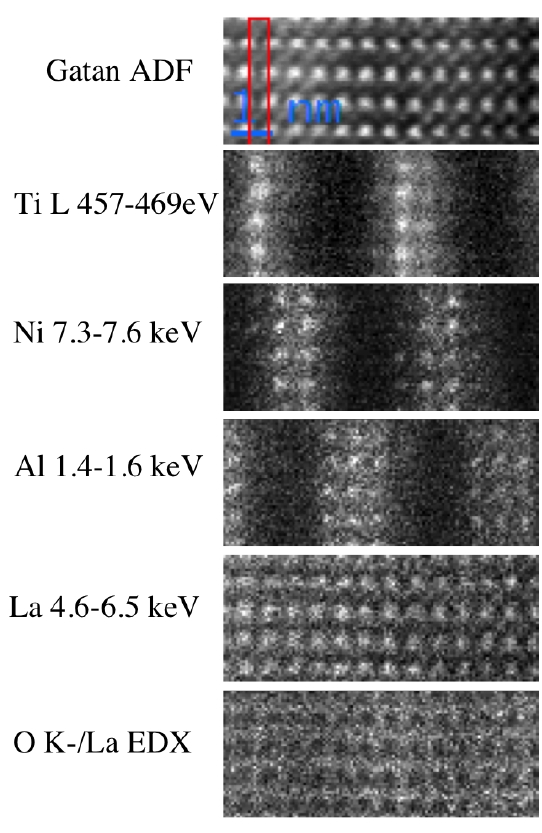}
\label{Fig:Spectroscopy}
}
\subfigure[]
{
\includegraphics[width=0.465\columnwidth]{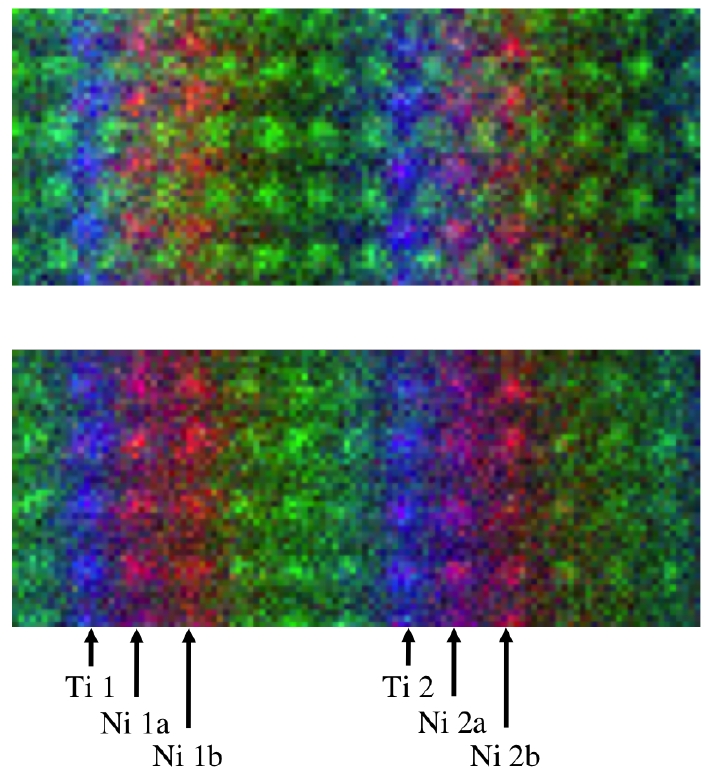}
\label{Fig:SI}
}
\caption{Spectroscopic analysis of the two superlattice repeats, shown in conjunction with a) all relevant chemical signals obtained either by integrating the EELS or EDX signal. b) Atomic-resolution spectrum images showing the three component superlattice. The top image shows the EELS signal, the bottom image is composed of the EDX signal. }

\end{figure}

The remaining figures present the relevant EELS fine structure results, beginning with the \TL\ (Figure \ref{Fig:Ti-EELS}), integrated from the indicated Ti column. The clear presence of four peaks is the classic signature of \Ti, resulting from the splitting of the degenerate $3d$ final states into the \Tg\ and \Eg\ levels for each of the Ti \Ltw\ and \Lth\ edges.\cite{Kourkoutis10} This is in contrast to bulk LTO, where the Ti is in a 3+ ($d^1$) state which leads to less well-defined \Tg-\Eg splitting, and a markedly different \TL\ signature all of which is easily identified via EELS.\cite{Ohtomo02} That the Ti in the LTNAO superlattice is 4+ is the first piece of direct evidence of the desired donation of an $e^-$ from Ti. The 4+ state of Ti in the superlattice is consistent with the previously reported XAS data, which averages over the entire superlattice film (Figure \ref{Fig:Ti-XAS}); XAS has the advantage of a considerably higher energy resolution than EELS but provides little spatial resolution. Analysis of the \NL\ to describe the Ni valence in the LNO layers is not possible here due to the nearly complete overlap between the La $M$- and the \NL s. We want to further emphasize here that the observed change in Ti valence is not due to film stoichiometry, i.e. oxygen vacancies. Figure \ref{Fig:Spectroscopy} shows the integrated \OK\ intensity normalized to the La concentration for the superlattice. It can be seen that the oxygen stoichiometry for all three layers does not vary, and the EEL spectra from the \LAO\ layers (which can be considered as a bulk reference in this context) show the fine structure expected for stoichiometry \LAO. Therefore, it appears that all layers in the superlattice are stoichmetric and the observed changes in the valence and EELS fine-structure are associated with interfacial charge transfer.
In the following, we will therefore focus on the \OK\ analysis to extract and quantify the interfacial charge transfer.

\begin{figure}[h]
\subfigure[]
{
\includegraphics[width=0.465\columnwidth]{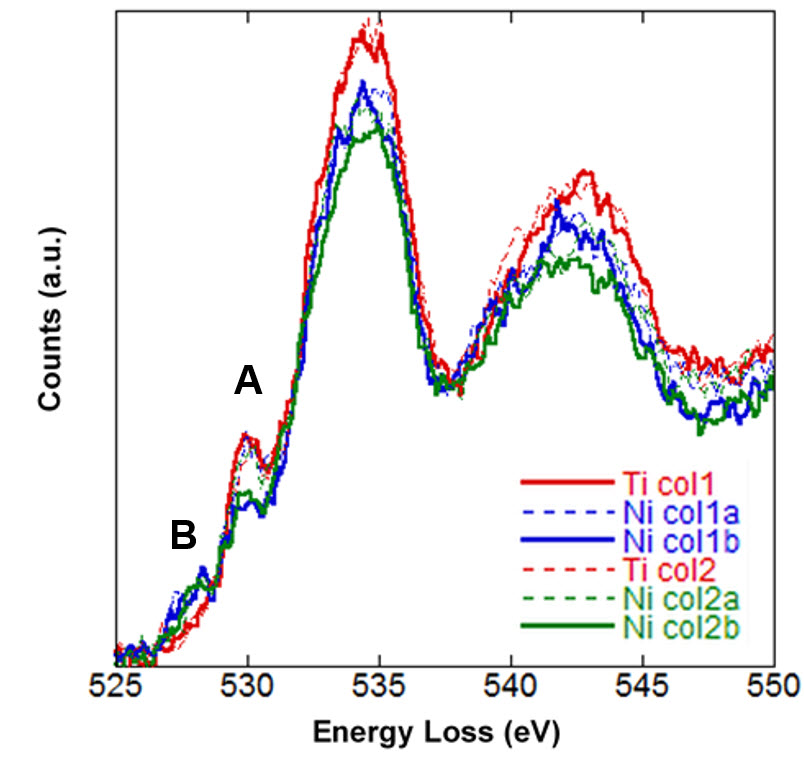}
\label{Fig:Ok-EELS}
}
\subfigure[]
{
\includegraphics[width=0.465\columnwidth]{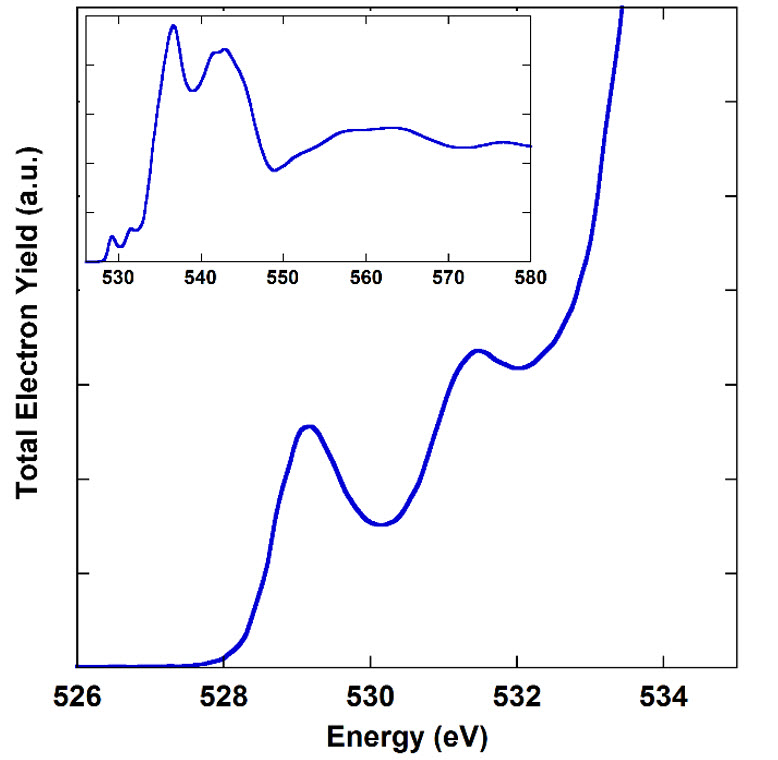}
\label{Fig:Ok-XAS}
}
\subfigure[]
{
\includegraphics[width=0.465\columnwidth]{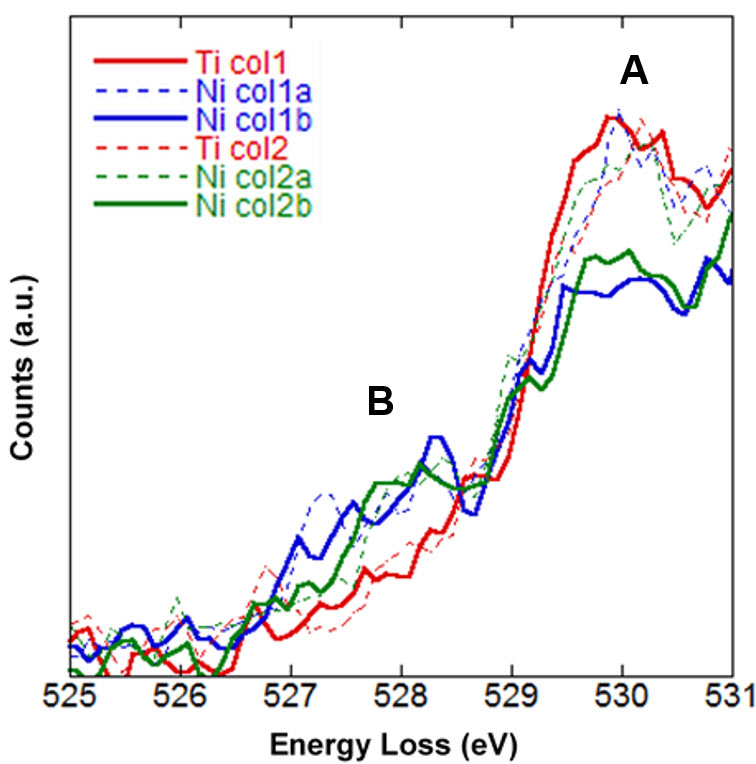}
\label{Fig:Ok-Zoom}
}
\subfigure[]
{
\includegraphics[width=0.465\columnwidth]{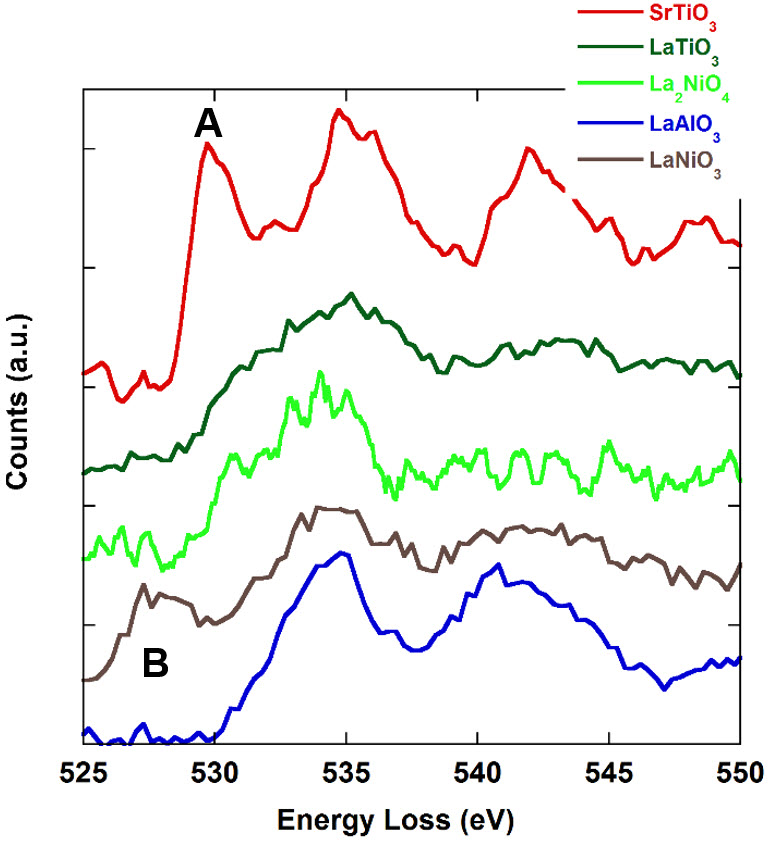}
\label{Fig:Ok-ref}
}

\caption{ \OK\ spectra: a)  EEL spectra integrated (vertically) over the atomic columns indicated in Figure \ref{Fig:SI}. b) X-ray absorption spectrum of the entire superlattice film; the inset shows an overview of a larger energy range. c) magnified pre-peak region shown in a) for the \LTO and \LNO layers with the different pre-peaks labeled $A$ and $B$. d) Reference spectra for Ti$^{4+}$ (\STO),  \Ti\ (\LTO), \Ni\ (\LNO), Ni$^{2+}$ (\LaNiO) and \LAO. }
\label{Fig:EELS-Fine-structure}
\end{figure}

Various \OK\ spectra are provided for the Ti and Ni columns contained within the spectroscopic region of interest, again integrating a number of  rows, along the respective columns (Figure \ref{Fig:EELS-Fine-structure}). While there are some difference in the absolute intensity of the main peak, we will be focusing on the pre-peak of the \OK, which results from electronic transitions into the hybridized O $2p$ - transition metal (TM) $3d$ orbitals.\cite{Abate92} What is visible in all the spectra is a pre-peak centered around 530 eV (labeled $A$), and an additional peak near 528 eV ($B$) which is only present in the spectra from the Ni columns. These peaks are readily explained by comparison to known bulk EEL spectra, shown in Figure \ref{Fig:Ok-ref}. For example, looking at bulk LTO (TM valence is 3+), we see merely a slight pre-shoulder on the main peak, due to its $3d^1$ configuration, due to a decreased number of unoccupied states. In \STO, where the Ti $d$-orbitals are completely empty, a strong pre-peak intensity, $A$, is seen. Peak $A$ in the Ti column of the LTNAO superlattice is analogous to that seen in the STO as opposed to the LTO bulk reference, suggesting that Ti is in a 4+ valence state.

Examining a reference spectrum for bulk LNO (Figure \ref{Fig:Ok-ref}), we see a pre-peak at 528 eV, shifted lower in energy with respect to the LTO and STO pre-peaks located at an edge onset energy of 530 eV. In the LNO superlattice, the lower energy pre-peak is manifested as peak $B$ seen in the Ni columns, though of a reduced intensity compared to the bulk reference spectra for LNO, due to the acceptance of an e- from the Ti layer and hence a reduced Ni valence. Again, we expect a diminished pre-peak\cite{Suntivich14} in this case because the additional $e^-$ into the Ni layers reduces the number of empty hybridized O $2p$ - TM $3d$ orbitals probed by the incident electrons. Indeed, XAS of the \OK\ confirms the presence of both of these pre-peaks, $A$ and $B$. We reiterate that XAS is a spatially averaging spectroscopic technique and cannot tell us the specific spatial origin of these peaks.

Looking carefully at peak $A$ in all of the spectra shown in Figures \ref{Fig:Ok-EELS} and (c), it is apparent that there is still significant spectral weight in both Ni columns. Given that there is some Ti present in the first Ni column of each superlattice (based on the EELS/EDX spectroscopy), some of this intensity in Peak $A$ can potentially be explained by remnant Ti contributions. However, when examining the \OK\ fine-structure of \LaNiO\ (Figure \ref{Fig:Ok-ref}), where the Ni valence state is expected to be 2+, we do not find any sign of the pre-peak $B$ at 528 eV (as seen for \Ni\ in \LNO\ reference), but instead a shoulder at $\approx$530 eV, which coincides with the position of peak $A$ in Figure \ref{Fig:Ok-Zoom}. The peak $A$ intensity in the spectra taken from the LNO layers is therefore not completely due to the remnant Ti contributions but also due to the contribution of Ni$^{2+}$.

It is interesting to note that the intensity of peak $A$ is significantly higher in the layer closest to LTO (i.e. Ni col1a and Ni col2a) and decreases in the layers closest to LAO (i.e. Ni col1b and Ni col2b). Without further insights from theoretical modeling, it is impossible to disentangle the contributions to this peak stemming from remnant \Ti\ in the LNO layer closest to LTO  and from the increasing Ni valence in the layer closest to the LAO layers. The authors acknowledge that with the electron probe in a channeling condition (i.e., a zone axis orientation) as is the case here, one must be cautious when attempting quantitative EELS and EDX measurements, as these experiments can be convoluted by elastic and thermal diffuse scattering of the incident electrons.\cite{Lugg12, Lugg14} However, in the case of the present fine structure analysis, we are simply looking at clear trends in the spectra, which appear and disappear rapidly, generally within a single unit cell, as shown in Figure \ref{Fig:Ok-EELS}; we furthermore note the good agreement the XAS data, which is insensitive to channeling.

For our DFT calculations, we simulated the LTNAO system strained to \LAO. In addition, we consider bulk NiO (with nominal Ni$^{2+}$ valence), bulk \LNO\ (nominally \Ni), bulk \LTO (nominally \Ti), and bulk \STO\ (nominally \Ti). We calculate the relaxed LDA atomic-scale structure as well as the orbital occupancies and the \OK spectra using both the $Z$ and $Z+1$ approximations.

The most useful comparison between theory and experiment comes from the calculated charge in the Ni $d$ orbitals, seen in Table \ref{Tab:1}. By comparing to the $d$ occupancy of NiO and \LNO, we determine an interpolated nominal charge on the Ni atoms in the two LNO layers of the LTNAO superlattices. This analysis gives a nominal charge Ni$^{2.3+}$ for the layer closest to LTO (i.e., Ni col1a and Ni col2a) and Ni$^{2.77+}$ for the layer closest to LAO (i.e., Ni col1b and Ni col2b). This result shows that the electron transfer from Ti is primarily limited to the LNO layers directly adjacent to LTO.

Due to the difference in charge, we would expect the \OK\ spectra to differ for the two Ni layers in the LTNAO. As a reminder, by comparison to bulk references (Figure \ref{Fig:Ok-ref}), we determine that the energy of pre-peak $A$ ($\approx$ 530 eV) primarily corresponds to Ti$^{4+/3+}$ states  plus  Ni$^{2+}$ states and the energy of pre-peak $B$ ($\approx$ 528 eV) corresponds primarily to \Ni\ states. Thus, from the calculated Ni charge, we would expect the intensity ratio of peaks $A$ and $B$ to be larger for the LNO adjacent to LTO (i.e., Ni col1a and Ni col2a)) than for the second LNO layer (Ni col1b and Ni col2b), reflecting the larger amount of Ni$^{2+}$ character. This prediction is in agreement with the experimental measurements and confirmed by integrating peak $A$ and $B$ intensities for the LTO, two Ni-a, and two Ni-b columns. The average $A/B$-peak ratio decreases as expected: 5.24 (Ti), 3.26 (Ni-a), and 2.56 (Ni-b), demonstrating the larger amount of Ni$^{2+}$ character in the Ni column adjacent to LTO.

It is worth noting that, despite much effort, a direct comparison of EELS data to theoretically calculated \OK\ spectra did not produce good agreement of the energies or intensities of the peaks for bulk LNO or LTNAO. Hence we do not rely on them in our theoretical analysis. Given that DFT is a ground state theory, it can, in principle, correctly compute the mean electron density and orbital occupations. However, as is well known, using DFT to predict electronic excitations (such as EELS) is much more problematic. Furthermore, localized dynamical electronic correlations are not included in DFT band structures, further degrading comparisons to experiments in such correlated complex oxides. We have, therefore, focused on using the observables that should be predicted correctly by DFT: the electronic density and mean occupancy of orbitals.

\section{Conclusions}

In summary, by combining atomically-resolved energy-loss and X-ray data with first principles DFT calculations, direct evidence is provided of the charge transfer from LTO into LNO in tricomponent superlattices. Using the high spatial sensitivity of STEM imaging and electron spectroscopies, we confirm previous XAS measurements, which have reported a $\approx$50 \% change in the orbital occupation that is significantly higher (by a factor of 2-3) compared to previous results.\cite{Wu13} Furthermore, we demonstrate that this interfacial charge transfer from the LTO to the LNO layers is highly localized in real space and is limited to the layers directly adjacent to each other. The range of the interfacial charge transfer is of the order of 1 unit-cell or about 4 \AA. These types of results and analysis provide crucial feedback for future orbitally-selective synthesis methods, where the magnitude and  range of charge transfer and orbital polarization will be used to stabilize novel electronic phases inaccessible by conventional epitaxial methods.

\section*{Acknowledgements}
PJP and RFK acknowledge funding from the National Science Foundation (NSF) via Grant Number DMR-1408427 to support this work. Work at Yale supported by NSF MRSEC DMR-1119826 (CRISP) and AFOSR under grant number FA9550­15­1­0472. Use of the National Synchrotron Light Source at Brookhaven National Lab was supported by the Office of Science, Office of Basic Energy Sciences, of the US Department of Energy under Contract No. DE-AC02-98CH10886.

\newpage


\newpage

\newpage

\newpage

\newpage

\newpage

\begin{table}[h!]
\begin{tabular}{c|c|c|c} \hline \hline
 Structure & Site  & $d$ electron occupancy & Hole ratio, r \\ \hline
 NiO (Ni$^{2+})$ & & 8.76 & \\ \hline
 LaNiO$_3$ (Ni$^{3+}$) & & 8.64 & \\ \hline
 LTNAO (LaAlO$_3$ strain) & Ni-a & 8.72 (2.33+) & 0.60 \\ \hline
 & Ni-b& 8.67 (2.77+) & 0.88 \\ \hline \hline

\end{tabular}
\caption{DFT calculated Ni $d$ occupation and hole ratio $r = h_{3z^2-r^2}/h_{x^2-y^2}$ for nickelate references and superlattices ($h_i$ indicates the hole occupation of orbital $i$).}
\label{Tab:1}
\end{table}

\end{document}